\documentclass[mathleft]{an}
\usepackage{graphicx}
\usepackage{color}
\usepackage{times}
\begin{document}
\Pagespan{789}{}
\Yearpublication{2014}%
\Yearsubmission{2014}%
\Month{11}%
\Volume{999}%
\Issue{88}%

\title{Wide-Field Plate Archive of the University Observatory Jena\,\thanks{Based on observations obtained with telescopes of the University Observatory Jena, which is operated by the Astrophysical Institute of the Friedrich Schiller University.}}

\author{A.V. Poghosyan\inst{1}, W. Pfau\inst{1}, K.P. Tsvetkova\inst{2},
  M. Mugrauer\inst{1}, M.K. Tsvetkov\inst{2,3},
  V.V. Hambaryan\inst{1}\fnmsep\thanks{Corresponding author:
    \email{valeri.hambaryan@uni-jena.de}},
  \newline R. Neuh\"auser\inst{1}, \and D.G.Kalaglarsky\inst{2}}
\titlerunning{Jena Wide-Field Plate Archive}
  \authorrunning{A.V.Poghosyan} \institute{ Astrophysikalisches Institut und
    Universit\"ats-Sternwarte Jena, Schillerg\"a{\ss} chen 2, D-07745 Jena,
    Germany \and Institute of Mathematics and Informatics, Bulgarian Academy
    of Sciences, Sofia 1113, Bulgaria \and Institute of Astronomy, Bulgarian
    Academy of Sciences, Sofia 1784, Bulgaria}

\received{17 Feb 2014}
\accepted{19 Mar 2014}
\publonline{later}

\keywords{\emph{astronomical databases: catalogs}}

\abstract{ We present the archive of the wide-field plate observations obtained at the University Observatory Jena, which is stored at the Astrophysical Institute of the Friedrich Schiller University Jena. The archive contains plates taken in the period February 1963 to December 1982 with the 60/90/180-cm Schmidt telescope of the university observatory. A computer-readable version of the plate metadata catalogue (for 1257 plates), the logbooks, as well as the digitized Schmidt plates in low and high resolution are now accessible to the astronomical community.This paper describes the properties of the archive, as well as the processing procedure of all plates in detail.}

\maketitle

\section{Introduction}

The large astronomical plate collections provide unique resources for photometric and astrometric studies of astronomical objects. The plates may serve as a unique and crucial opportunity to investigate e.g. the long-term photometric behaviour of stars, identify pre-supernovae, get positional information of asteroids or comets, etc. . From this point of view preservation of the fragile glass plates as digital copies for quick and informative access by the astronomical community is a timely and important enterprise. In this paper we present the results of the archiving of the plate collection, obtained in the period 1963 to 1982 with the 60/90/180-cm  Schmidt camera of the University Observatory Jena, operated by the Astrophysical Institute of the Friedrich Schiller University (henceforward abbreviated as AIU). The archiving process comprises the setup of a plate inventory, the preparation of a computer-readable version of the plate metadata catalogue, the plate digitization with two different resolutions, as well as the astrometric solution of all images. At this time, this is the only fully digitized plate archive with astrometric solutions in Germany.

\section{Characteristics of the Schmidt telescope}\label{sec1}

\begin{figure*}
\resizebox{\hsize}{!}{\includegraphics[angle=0]{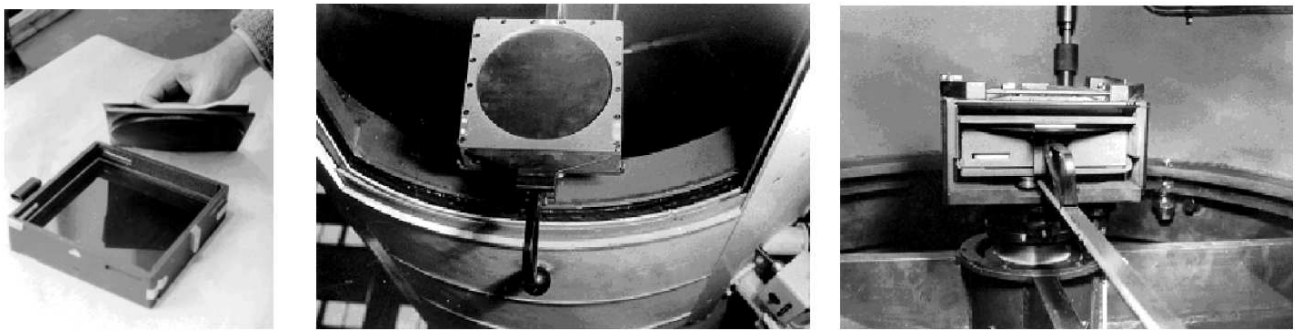}}
\caption{\emph{Left:} The plate cassette, consisting of the plate holder and the thrust plate, used to correct for the spherical field curvature of the Schmidt camera. \emph{Middle:} The plate cassette in its loading position at the outer tube of the telescope is shown. \emph{Right:} The plate cassette finally placed in the focus of the Schmidt camera.}
\label{Fig_focusstation}
\end{figure*}

The plates from the collection presented here were taken with the 60/90/180-cm Schmidt camera, which was installed at the University Observatory Jena in the end of 1962. The observatory is located close to the small village Gro{\ss}schwabhausen, about 10 km to the west of Jena. Architect of the telescope building was Hans Schlag, an architect who together with J. Schreiter is also famous for the Planetarium in Jena, built in 1926. The building is now on the monument heritage list. The telescope with a fork mount was primarily designed by Carl Zeiss Jena to be operated as a Schmidt camera of 60 cm free aperture and a diameter of the spherical main mirror of 90 cm. It exhibits a focal length of 180 cm. There is no field flattening lens and the plates of size 16\,cm\,$\times$\,16\,cm are bent within a cassette to fit to the spherical focal surface. As a second optical version a quasi-Cassegrain system with Nasmyth focus of 13.5 m focal length is provided. Restricted to the optical axis region it allows photometric and spectrometric work (Marx \& Pfau 1965, 1966a). In addition, two auxiliary telescopes are fixed to the main tube, a 20-cm refractor and a 25-cm Cassegrain system. Nowadays, all these telescopes are equipped with state-of-the-art CCD cameras (Mugrauer 2009; Mugrauer \& Bertholdt 2010) and fibre-fed spectrographs (Mugrauer \& Avila 2009, Mugrauer, Avila \& Guirao 2014) are in operation at the Nasmyth focus of the 90-cm telescope. All instruments as well as the telescope are operated from a control room in the first floor of the building.\\ The plate metadata catalogue of the Wide-Field Plate Archive of the University Observatory Jena, as presented here, is a part of the Wide-Field Plate Database\footnote{see http://www.wfpdb.org} (henceforward abbreviated as WFPDB). This is a unique source for plate metadata information and quick plate visualization. At the moment, the WFPDB includes plate metadata for about 600\,000 plates, equal to about 24\% of all wide-field plates stored worldwide, observing logbooks and other paper files affiliated to them, e.g. images of the original plate envelopes, plate quick visualization scans, some high resolution scans, linkage between some of the archival plates and the published papers based on this material, graphical representations of some plate data, etc. .

The 1997 version of the WFPDB is also accessible at the catalogue service of the Centre de Donn\'ees  de astronomiques de Strasbourg (CDS) as VizieR catalogue VI/90\footnote[2]{see http://cdsarc.u-strasbg.fr/viz-bin/Cat?VI/90}. Extraction of plate metadata from the whole database is possible via search masks. Part of the WFPDB is the Catalogue of Wide-Field Plate Archives (CWFPA). Information on the AIU plate archive is available as summarized in Table~\ref{Tab_PAinfo}.

\begin{table*}[h]
\caption{General information on the AIU plate archive, which became part of the WFPDB.}
\label{Tab_PAinfo}
\begin{tabular}{ll}
\hline
Observatory identifier in the WFPDB   &        JEN                          \\
Instrument identifier in the WFPDB    &        060                          \\
Location of the archive		      &  AIU   \\
Observatory			      &	 University Observatory Jena\\
Observatory code $^1)$		      &  134                                \\
Observatory coordinates               &  $\mathrm{E}\,11\degr \,29\farcm0\,; \ \mathrm{N}\,50\degr \,55\farcm8 $   \\
Altitude	                      &  356 m above sea level              \\
Type of telescope                     &  Schmidt camera                     \\
Clear aperture                        &  0.60 m                             \\
Mirror diameter		              &  0.90 m	                            \\
Focal length 			      &  1.80 m	                            \\
Size of circular field                &  $5\fdg1$                           \\
Image scale			      &  $115\,\arcsec$/mm                  \\	
Years of operation		      &	 1963 -- 1982                       \\
Number of plates		      &	1257                                \\
\hline
\end{tabular}

$^1)$ According to IAU Minor Planet Center (see http://www.minorplanetcenter.net/)\\
\end{table*}

The first Schmidt camera plate was obtained on December 2, 1962. Science observations with direct imaging started in the beginning of 1963. Later on, the Schmidt camera was also used together with a $1\degr$- objective prism for low-resolution spectroscopy as well (Marx, Pfau \& Richter 1971). Between 1963 and 1982 more than 1250 astronomical photographic plates had been taken. At present, in the plate collection of the AIU 1159 plates are available ($\approx 92\%$), the remaining 98 plates are missing and may still be kept unwittingly by colleagues. Over the years quite a number of people was involved in the observations with the Schmidt camera. The majority of all plates were taken by three observers, namely W. Pfau, S. Marx, and D. Uhlig, who obtained about 851 plates, i.e. 68\% of all plates.\\ In preparation of an exposure, the plates had do be inserted into a cassette and moved to the focus position with the aid of a carriage. Within the cassette, the glass plate was pressed from its backward side against a circular ring and thus adjusted to the spherical focal surface with a curvature radius of 1.8 m (see Fig.~\ref{Fig_focusstation}).

\section{Preparation of the plate catalogue}

The main source of information used for the preparation of the computer-readable plate catalogue were the telescope logbooks. The catalogue is compiled in the WFPDB format and therefore contain information on the equatorial J2000.0 coordinates of the plate centre, the date and starting time of the observation (in UT), the object name and type, the method of observation,  the number of exposures and their duration,  the type of  the used emulsion, filter and colour band,  the size and quality of the plate, notes with the name of the observer(s), the place of the plate storage (availability), as well as the status of plate digitization. The original logbook of the 60/90/180-cm Schmidt telescope and the observers notes were scanned and a link of the metadata information to the logbook page, showing the entries for the relevant plate, is provided online on the WFPDB website.  A search in the \mbox{WFPDB}, using JEN\,060 as observatory and telescope identifier,  can be done either by object or field coordinates, or by constraints on the observational parameters. In addition to the information retrieved, the user can display an additional page with details for the selected JEN\,060 plate (see Section~\ref{sec1}).

\section{Plate catalogue analysis}

\begin{figure*}
\resizebox{\hsize}{!}{\includegraphics[angle=0]{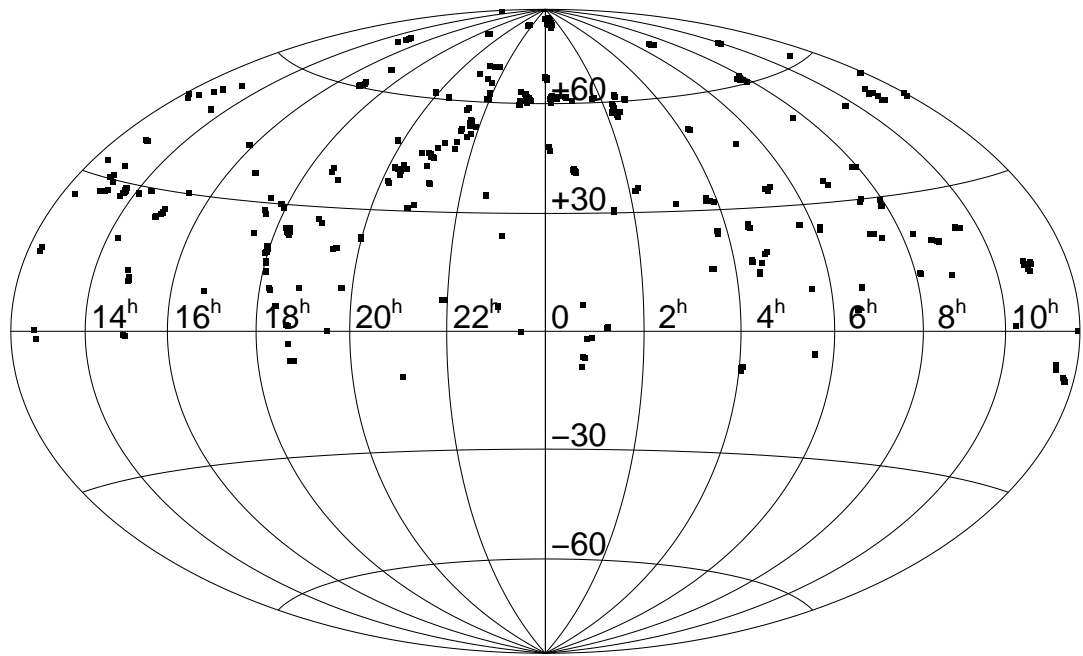}}
\caption{The centres of all digitized Schmidt plates in equatorial coordinates in Aitoff projection.}
\label{Fig_map}
\end{figure*}

The preparation of a computer-readable version of the JEN\,060 plate catalogue and its inclusion into the WFPDB in the standardized format was done to achieve general accessibility and re-use of the photographic plates by the whole astronomical community. The evaluation of the catalogue as given below refers to essential items, such as the all-sky distribution of plate centres, object types, distribution of epochs of
observations, duration and multiplicity of exposures, emulsions and use of glass filters, in order to match certain broad-band photometric systems. The plate catalogue is accessible online\footnotemark[1]. It comprises metadata information for 1257 plates, obtained in the period February 1963 to December 1982. A map of the sky coverage of all plate centres in equatorial coordinates is shown in Fig.~\ref{Fig_map}. 

\begin{figure}
\resizebox{\hsize}{!}{\includegraphics[angle=-90]{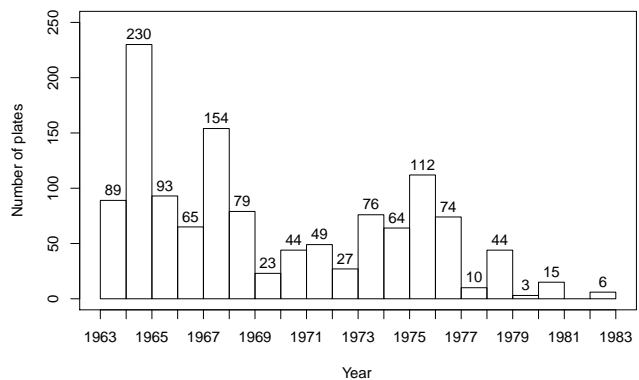}}
\caption{Number of plates taken by year.}
\label{Fig_yearly}
\end{figure}

\begin{figure}[h!]
\resizebox{\hsize}{!}{\includegraphics[angle=-90]{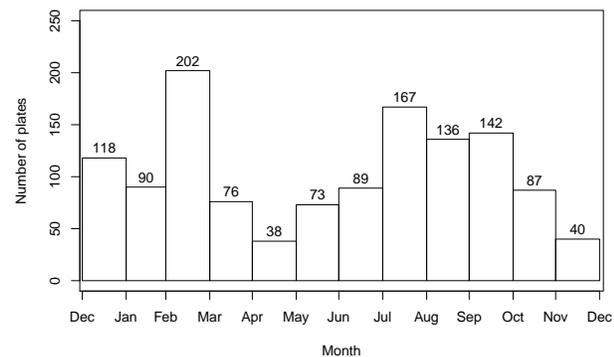}}
\caption{Number of plates taken by month (sum over the years).}
\label{Fig_monthly}
\end{figure}

The distribution of plates over time is shown in Fig.~\ref{Fig_yearly}
and \ref{Fig_monthly}. There are three maxima of observing activity, namely the years of 1964, 1967, and 1975, with declining activity afterwards. The decline reflects the continually increasing use of the telescope's Nasmyth focus for photoelectric observations. The cessation of imaging in the Schmidt focus in 1982 was mainly caused by the growing difficulties to acquire photographic plates for astronomical use from ORWO, the plate manufacturer in East Germany. The distribution by month mainly reflects the typical seasonal weather statistics at the location of the University Observatory Jena.

\subsection{Telescope tests \label{ttests}}

Because of the close relations between the AIU and Carl Zeiss Jena, thorough and intensive technical tests of the Schmidt camera were performed. Among others, they referred to modified Hartmann tests of optical aberrations and light concentration (Marx \& Pfau 1966a), tests of the focus stability under changing temperature and other influences, tests of the stability of the relative orientation between the optical axes of the guiding telescope and the Schmidt camera, tests of the bending of the fork-mount with the telescope position from varying gravitational load, and tests of the periodic errors of the telescope tracking. Different experiments with emulsions and filters were done and limiting magnitudes derived. Most of the tests were performed on photographic plates. Of special importance was the successful test of a $1\degr$-objective prism of new design. In order to reduce the number of optical parts at the entrance pupil
of the telescope to only one and to limit the overall thickness of the glass components in the beam, the manufacturer invented a demanding technique. The optical surface of the corrector-plate was ground directly to a wedge-shaped
glass plate. This single element successfully combines the effects of a Schmidt plate and that of a prism (Marx et al. 1971).

\subsection{Science observations \label{sciobs}}

From the beginning, the telescope was used for different programmes and a diversity of object types. 82\% of all 1257 plates were used for direct imaging observations. Multi-exposures on about 100 plates were applied in case of observations of minor planets and comets, sometimes on open clusters, supernovae, {\sc H\,ii} ionized regions, and extremely red stars. 181 plates
(14\% of all) were taken for technical purposes.\\ The observing programmes, together with the relevant numbers of plates, are presented in Table~\ref{Tab_SciProgr}. Quite often, plates taken within the frame of a certain programme were used for another purpose, as well. For example, the plates from the supernova search programm also served the investigation of quasi-stellar objects (QSOs) or led as a by-product to the discovery of an eruptive variable star in Hercules (Dorschner \& Friedemann 1968).

\begin{center}
\begin{table*}[h]
\caption{Observing programmes executed with the AIU Schmidt telescope.}
\label{Tab_SciProgr}
\begin{tabular}{lcc}
\hline
Observing                           & Number         & Notes          \\
programme                           & of plates      &                \\
\hline
Supernova Search                    & 180            & 1              \\
Optical behavior of QSOs	    & 41             & 2              \\
Galaxies                            & 66             & 3              \\
Open stellar clusters               & 221            & 4              \\
Globular clusters                   & 43             & 5              \\
Stellar associations                & 81             & 6              \\
Stars in general                    & 77             & 7              \\
Variable stars                      & 87             & 8              \\
Extremely red stars and Jena Variables (JV)
                                    & 44             & 9              \\
Planetary nebulae                   & 4              & 10             \\
H II ionized regions                & 76             & 11             \\
Comets 	                            & 44             & 12             \\
Planets and Minor planets           & 5              & 13             \\
Radio sources                       & 5              & 14             \\
Stellar Fields	                    & 9              & 15             \\
Interstellar reddening	            & 18             & 16             \\
Joint observations with the         & 11             & 17             \\
astronomical institutes             &                &                \\
\hspace{0.2cm} at Sonneberg, Babelsberg, and Tautenburg
                                    &                &                \\
Technical telescope tests           &  181
                                    &  see subsection \ref{ttests}    \\
\hline
\end{tabular}

{\bf Notes:}\\ $[1]$ \ \ see subsection \ref{SNe} \\ $[2]$ \ \ 3C 245, 3C
309.1, 3C 351, 3C 390.3, 3C 435.1, PKS 0106+01 (4C 01.02), PKS 0115+02 (3C
37), QSOs around BD +33 2298, and BD +39 2580. The observations were done in
the period 1966 to 1978, part of the results were published by Pfau (1973)
\\ $[3]$ \ \ Seyfert galaxies Mrk 10, Mrk 6, M51, the Andromeda Galaxy M31,
the Triangulum Galaxy M33, the group of M81 and M82, the LINER-type Active
Galaxy Nucleus of M106, NGC 7331 \\ $[4]$ \ \ The Pleiades, Praesepe, Hyades,
h and $\chi$ Per, Coma Berenices cluster, NGC 103/NGC 146, NGC 188, NGC 225,
NGC 663, NGC 1528, NGC 1647, NGC 2158, NGC 2237/39, NGC 6709, NGC 6939, NGC
7023, NGC 7086, NGC 7128, NGC 7142, NGC 7160, NGC 7654 (Pfau 1980), IC 1311, IC
1805, M39, and King 13 (Marx \& Lehmann 1979) \\ $[5]$ \ \ M3, M13, and NGC
5466 \\ $[6]$ \ \ Orion nebula, North America nebula, and the structure of
interstellar matter around $\zeta$\,Persei (K\"uhn 1973) \\ $[7]$ \ \ $\zeta$
Dra, 63 Tau, HD 173850, and HD 228244; binary systems $\Theta$ Gem, $\delta$
Gem, $\gamma$ Aql, 83 Tau, and $\varepsilon$ Sct; Blue supergiant star AE And,
high-proper motion star $\chi$ Her; Be-stars CW Cep, MWC 645, and $\gamma$
Cas; BD stars BD+25 0718, BD+28 2185 (or 41 Com), BD +33 2235, BD +33 2298, BD
+36 4458, BD +61 1939; Carbon star IRAS 18400-0704 \\ $[8]$ \ \ Different
types of variability were observed: $\alpha$ Per, $\alpha$ Peg, $\gamma$ Cyg,
Nova Cygni; Orion type star AE Aur; Pulsating variable stars $\alpha$ Cyg and
$\alpha$ Tau; Semi-regular pulsating star RU Cyg; RR Lyr type star DD Lyr; the
prototype of the variable stars of $\alpha ^2$\,CVn; $\delta$ Sct type stars
as $\alpha$ Lyr and EH Lib; Spectroscopic binary $\eta$ CrB(1); the RS CVn
type variable $\xi$ UMa; Eclipsing binary of Algol type (detached) $\beta$
Aur; $\beta$ Cep \\ $[9]$ \ \ Extremely red stars, earlier discovered at AIU
on POSS charts were investigated (Friedemann, G\"urtler \& Pfau
1977). Besides, this resulted in the detection of 15 new red Jena Variables
(JV) in the Cepheus and Cygnus regions (Pfau \& Friedemann 1980) \\ $[10]$
\ M57 and NGC 6853 \\ $[11]$ \ IC 1848, NGC 7822, and NGC 7635 \\ $[12]$
\ These ''targets of opportunity'' are collected in Table~\ref{Tab_comets},
some of these being published in Marx \& Pfau (1967a) \\ $[13]$ \ Planet
Saturn and the minor planets Endymion (342), Alma (390), Asteria (658),
Maritima (912), Herluga (923), Jeanne (1281), Zamenhof (1462), Yugoslavia
(1554), and van Gent (1666) \\ $[14]$ \ 5C 2  \\ $[15]$ \ Standard
stars in Selected Area SA 61, fields around $\delta$ Lyr, in Cygnus, and in
the Milky Way \\ $[16]$ \ Field around $\omega$ Boo to investigate the
interstellar extinction at high galactic latitudes. The programme was
supplemented by photoelectric observations in the Nasmyth focus of the
Gro{\ss}schwabhausen telescope (Pfau 1979)\\ $[17]$ \ Some of the long-exposure
plates were jointly taken with colleagues from Zentralinstitut f\"ur
Astrophysik, Sternwarte Babelsberg on objects from the 5C catalogue of radio
sources \\ $[18]$ \ see Subsection 4.1\\
\end{table*}
\end{center}

\begin{center}
\begin{table*}[h]
\caption{Observations of comets}
\label{Tab_comets}
\begin{tabular}{lc}
\hline
Name of comet                       &  Number      \\
                                    & of plates    \\
\hline
Kilston (C/1966 P1) 	            &        16    \\
Kohoutek (C/1973 E1)		    &	      7    \\
Barbon (C/1966 P2)		    &	      6    \\
Alcock (C/1965 S2) 		    &	      5    \\
Kobayashi-Berger-Milon (C/1975 N1)  &         5    \\
Tsuchinshan 2 (P/1965 A1)           &         4    \\
Shajn-Schaldach (P/1949 S1)         &         1    \\
\hline
\end{tabular}
\end{table*}
\end{center}

\subsubsection{Supernova search \label{SNe}}

During the period June 1964 to April 1976 a monitoring programme was executed at the AIU in order to detect supernovae in selected galaxy fields. As a culmination of his work on supernovae, F. Zwicky in 1961 proposed to the General Assembly of the IAU that research in the field should be done in an international campaign, coordinated by IAU Commission 28 (Galaxies). As one of these tasks the brighter galaxies should be watched at several observatories world-wide with medium-sized Schmidt telescopes (Zwicky 1964). At the AIU most of the plates in
the supernova programme were taken up to the end of 1968. The list of the patrol fields and number of plates taken is given in Table~\ref{Tab_SN}, exposure times being mainly 20 or 30 min. On plates taken in SN Field 17, the brightness of a supernova in NGC 3389 could be monitored for two and a half months (Marx \& Pfau 1967b). In total, 178 plates were taken at the AIU in the search for supernovae. Two other plates, imaging the Cygnus Loop are also counted within this programme.\\

\begin{center}
\begin{table*}[h]
\caption{Supernova searches executed with the AIU Schmidt telescope.}
\label{Tab_SN}
\begin{tabular}{cccc}
\hline
Field number          &	R.A. (2000.0)	&  DEC (2000.0)   &  Number    \\
                      &  ( h m s)       &  ($^{\degr}$ ' '') & of plates  \\
\hline
1                     &  00 \ 06 \ 37   & +47 \ 52 \ 38   & 	 8     \\
2, NGC 1647           &  01 \ 15 \ 34	& +00 \ 58 \ 58   &	16     \\
3		      &  02 \ 13 \ 16	& +36 \ 16 \ 23   &	20     \\
4		      &  08 \ 16 \ 53	& +21 \ 07 \ 00	  &	 3     \\
5                     &  07 \ 57 \ 20	& +29 \ 22 \ 02	  &	 4     \\
9		      &  14 \ 24 \ 08	& +29 \ 18 \ 00   &	 1     \\
10                    &  17 \ 52 \ 04	& +30 \ 05 \ 14	  &	10     \\
11, NGC 680           &  19 \ 31 \ 00	& +20 \ 16 \ 00   &	 4     \\
11                    &  18 \ 15 \ 47	& +23 \ 59 \ 33   &	56     \\
12                    &  19 \ 50 \ 47	& +48 \ 10 \ 17	  &	 1     \\
13                    &  14 \ 54 \ 33	& +11 \ 18 \ 31   & 	 8     \\
14                    &  12 \ 20 \ 34	& +14 \ 44 \ 20	  &	 2     \\
15		      &  11 \ 57 \ 02	& +47 \ 47 \ 47	  &	 3     \\
17                    &  10 \ 46 \ 43	& +12 \ 55 \ 06	  &	25     \\
18		      &  11 \ 52 \ 21	& +53 \ 42 \ 00   & 	 1     \\
20	 	      &  16 \ 54 \ 47	& +60 \ 25 \ 34	  &     10     \\
SN Gates in NGC 2903  &  09 \ 32 \ 09	& +21 \ 3  \ 02	  &      2     \\
SN Chavira	      &  12 \ 41 \ 40   & -01 \ 26 \ 58   & 	 4     \\
\hline
\end{tabular}
\end{table*}
\end{center}

\subsection{Photographic emulsions and photometric systems}

As specified in the logbooks, during
the first years of telescope operation, Agfa emulsions (Agfa Astro Spezial, Agfa Press Panchromatic, Agfa Spectral, Agfa Spectral Rot Rapid, Agfa Raman) were used, later on then emulsions from ORWO (ZP\,1, a few ZP\,2,  ZP\,3, and ZU\,2). Kodak plates were not available to the AIU at that time. In publications by Breido (1967) for the panchromatic ZP\,1 and ZP\,3 plates and by H\"ogner \& Ziener
(1978) for the blue-sensitive ZU\,2, comparability to the Kodak plates is demonstrated. According to these authors, in the red the ZP\,3 can entirely replace the 103a-E. Working in the blue band, the ZU\,2 shows a somewhat higher sensitivity than the Kodak emulsions 103a-O and IIa-O. To represent certain broad band photometric system, the combinations of ORWO emulsions with Schott filters are given in Table~\ref{Tab_UBV}. 967 plates (77\% of all) were taken in the Johnson broad band photometric UBVR-system. The statistics of exposure times shows up to 30 minutes for 76\% of all 1076 science plates and 60 minutes or more for the remaining 24\%. Long exposures (above 90 minutes) were applied for observations of the {\sc H\,ii} ionized region NGC 7635, the Babelsberg programme (120-150 minutes), the peculiar galaxy M106, the globular cluster M3, as well as the North America nebula.

\begin{center} 
\begin{table*}[h]
\caption{Broad band photometric systems and limiting magnitudes \newline at
  the 60/90/180-cm telescope (Marx \& Pfau 1966b)}
\label{Tab_UBV}
\begin{tabular}{clc}
\hline
Band     & Emulsion \& Schott filter & limiting mag. $^2)$    \\    
         & combination $^1)$         & (expos. time in min.)  \\
\hline
pg $^3)$ & Agfa Astro Spezial,       &                        \\
         & Agfa Press or             &                        \\
         & ZU2, without filter       &                        \\
U $^4)$  & ZU2 + UG2 (2mm)           & 18.5 (60 min.)         \\
         & ZU2 + UG11 (2mm)          &                        \\
B $^4)$  & ZU2 + GG13 (2mm)          & 18.4 (25 min.)         \\
V $^4)$  & ZP1 + VG 10 (4 mm) + GG14 (1 mm)
                                     & 16.5 (60 min.)         \\
         & ZP1 + BG23 (1.2mm) + GG14 (3mm)     
                                     & 16.0 (60 min.)         \\
R $^5)$  & ZP3 + GG14 (2mm)          &                        \\
\hline
\end{tabular}

$^1)$ in brackets thickness of filter glass\\
$^2)$ Marx (1971)\\ $^3)$ old photographic magnitudes\\ $^4)$ defined by
H. L. Johnson \\ $^5)$ defined by G. E. Kron \& J. L. Smith \\
\end{table*}
\end{center}

\section{Plate digitization}

\begin{figure}[h]
\resizebox{\hsize}{!}{\includegraphics[angle=0]{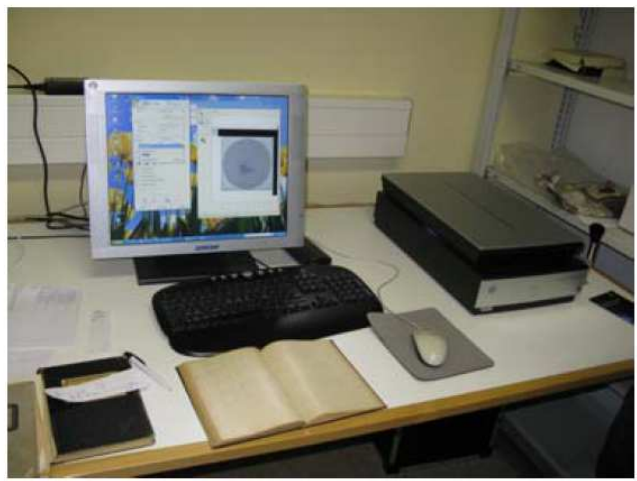}}
\caption{Office-lab for plate digitization at the AIU. The flatbed scanner, used for the digitization of all Schmidt plates, is shown on the right side.}
\label{Fig_Lab}
\end{figure}

In 2009, a office-lab for plate digitization was established at the AIU (Fig.~\ref{Fig_Lab}).

As a first step, the plates were placed in uniform and protecting envelopes. The process of digitization described here was primarily done to guarantee a lasting and reliable preservation of the information content, quick access to digital scan data, and thus facilitate plate re-using and sharing. The flatbed scanner Epson Perfection V700 Photo together with the software \emph{Scanfits}, developed by S. Mottola (see Barbieri et al. 2003) was used to scan all the available AIU Schmidt plates. The scanner combines high astrometric and photometric accuracy with fast scanning speed. The main parameters are summarized in Table~\ref{Tab_scanner}. In preparation of the scanning process, the glass surfaces were first cleaned from dust. Afterwards the plate previews were scanned, which still contain all marks drawn by the observers. For the final high-resolution scans of the plates all marks of the observers were then removed from the plates. The FITS header of the high-resolution scans contains the astrometric solution (see Sec.~\ref{astsol}) to the stellar field, done by the \emph{FTOOLS} software\footnote[3]{see  http://heasarc.gsfc.nasa.gov/ftools/} (Blackburn, 1995). Note that the keyword DATE-OBS in the FITS headers of the digitized files refers to UT. It was converted from the local time given in the logbooks, as well as the plate catalogue. Two scans of different resolution are available for any plate: a low-res JPEG file of 1200 dpi (file size per plate 2 MB) and a high-res FITS file of 2400 dpi (600 MB).

\begin{center}
\begin{table*}[h]
\caption{Technical parameters of the used scanner.}
\label{Tab_scanner}
\begin{tabular}{lc}
\hline
Type                         &   Epson Perfection Photo V700\\
                             &   flatbed scanner            \\
Maximum hardware resolution  &   $4800 \times 9600$         \\
(dpi)                        &                              \\
Maximum optical density      &   4.0                        \\
Colour depth (bits)          &   48/48                      \\
(internal/external)
                             &                              \\
Grayscale depth (bits)       &   16/16                      \\
(internal/external)
                             &                              \\
Maximum scan area (cm)       &   $21.5 \times 29.7$         \\
\hline
\end{tabular}
\end{table*}
\end{center}

The digitization process comprised three steps: (i) choice of scanning parameters and estimation of the quality of the digitized data, (ii) plate scanning with low resolution (1200 dpi) in the JPEG file format for quick plate visualization and quick online access (plate previews), (iii) working scans with optimal high resolution (2400 dpi equivalent to about 10 $\mu$m/pixel) in FITS file format to achieve photometric and astrometric quality, and (iv) linkage of the plate previews to the WFPDB for online access. The scans in their original file formats ($\sim$~1.7 TB) are stored on the AIU servers and are available upon request.

The 544 pages of the original AIU Schmidt telescope logbooks were also scanned. Logbook, observers notes, and the plate previews are accessible via search query in the WFPDB.

\section{Astrometric solution \label{astsol}}

Astrometric calibrations, i.e. transformation from image pixels into sky coordinates, have been derived for a total of 985 plates by means of tools provided by \emph{Astrometry.net}, an engine to transform any astronomical image coordinates into the world coordinate system with standards-compliant astrometric data.

\begin{figure}
\resizebox{\hsize}{!}{\includegraphics[angle=0]{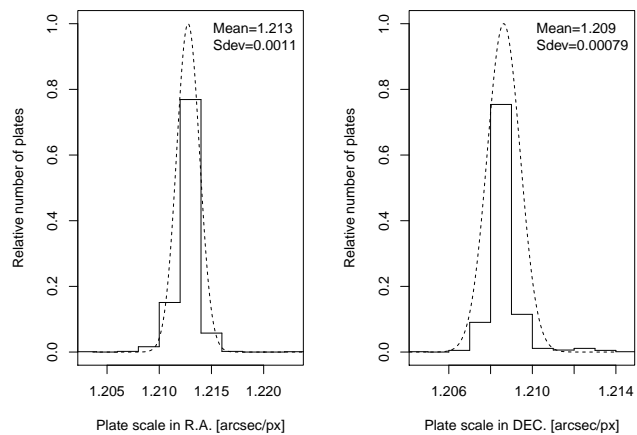}}
\caption{Histograms of plate scales (in arcsec/px) in Right Ascension (R.A.) and Declination (DEC.) directions for a total of 985 plates. The dashed lines show maximum-likelihood fitting of the corresponding plate scales with a Gaussian distribution.}
\label{Fig_scale}
\end{figure}

\begin{figure*}
\includegraphics[angle=0, height=98mm]{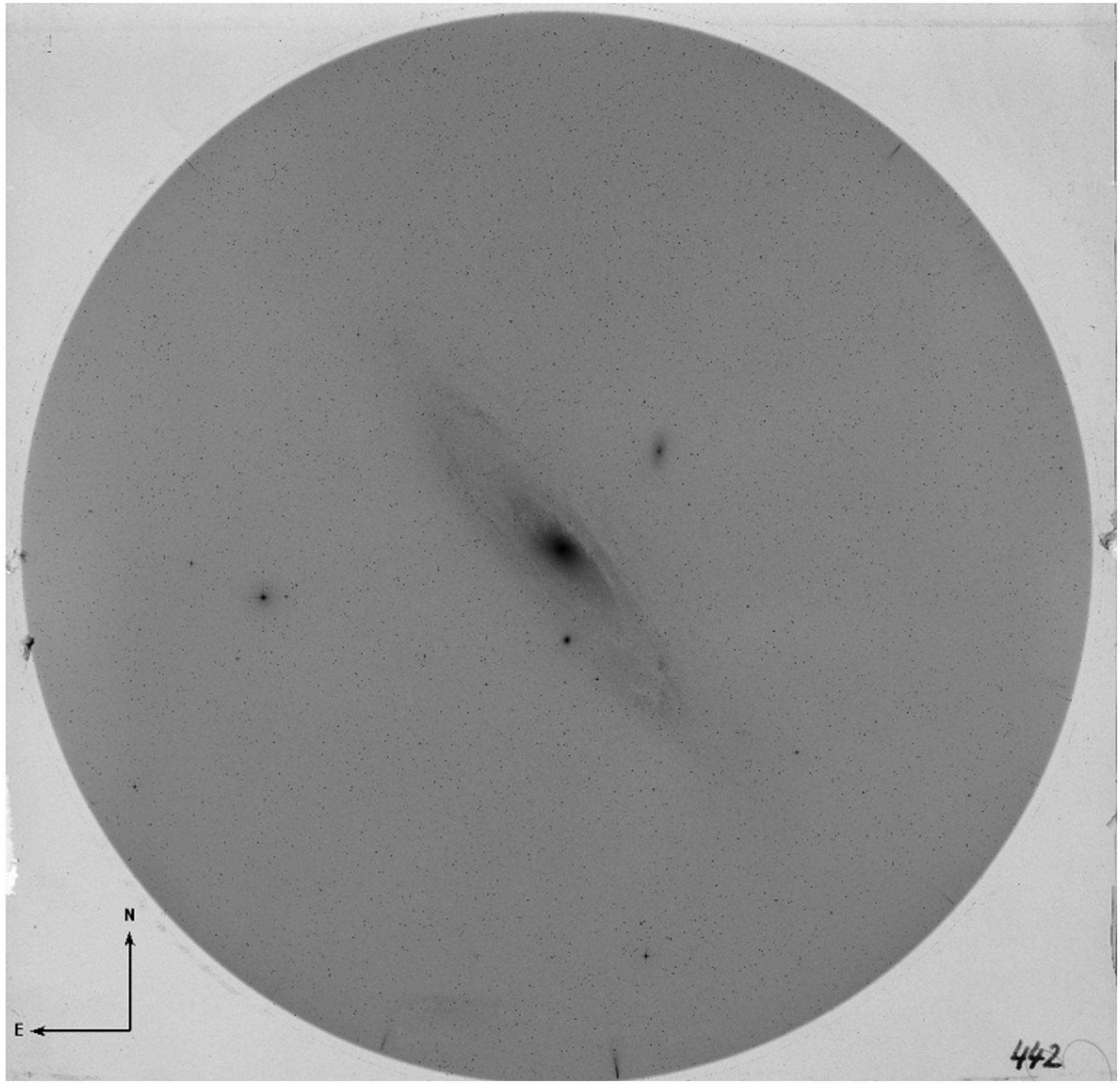}~~\includegraphics[angle=0, height=98mm]{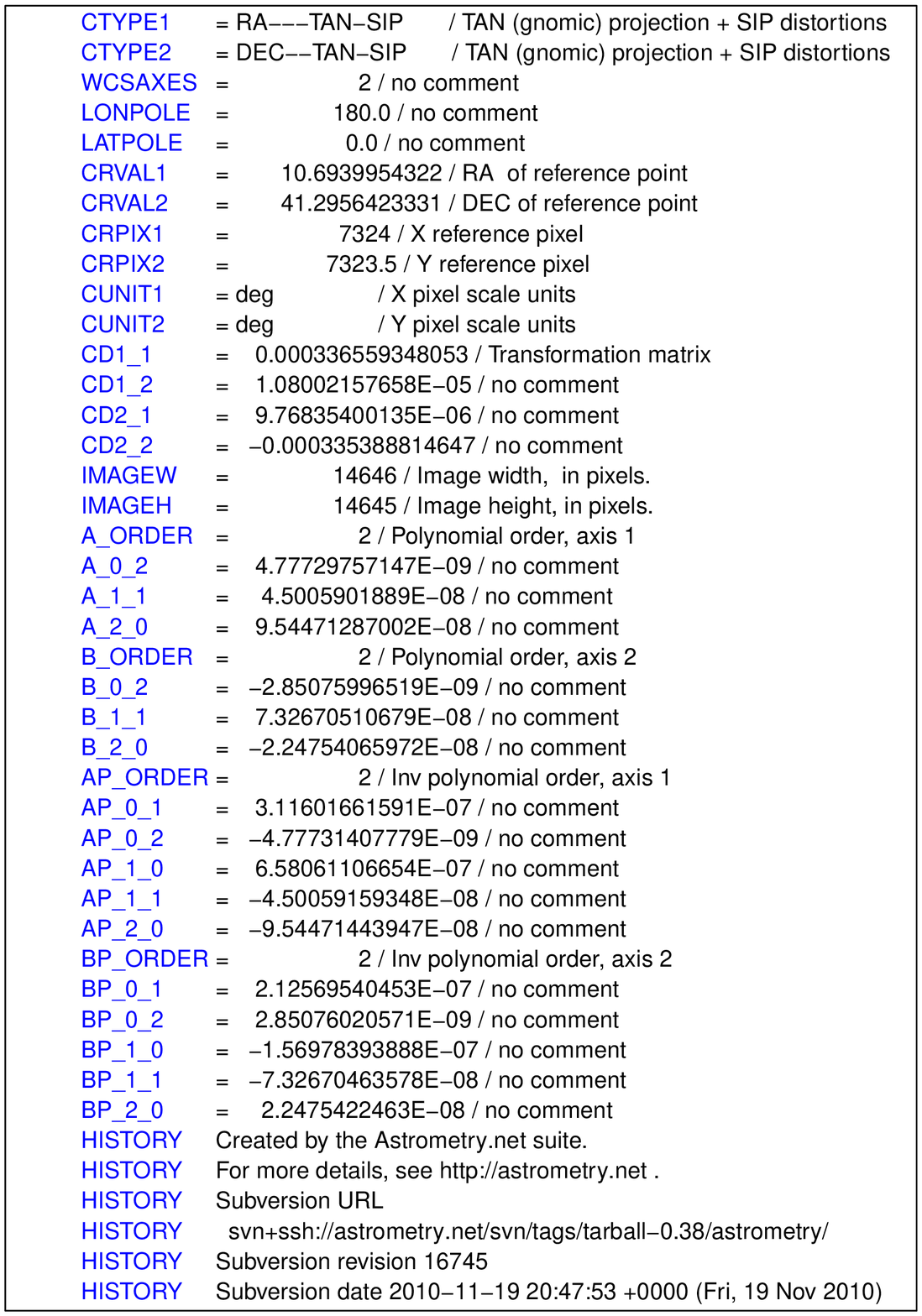} \caption{An example of a scanned Schmidt plate (JEN060 000442) of the Andromeda Galaxy (M31) and the portion of the header from the corresponding FITS file with its astrometric solution.}
\label{Fig_scan}
\end{figure*}

From the astrometric solutions of each image, the plate scales and position angles, were derived using \emph{extast} and \emph{getrot} routines from the IDL astro library (see Fig.~\ref{Fig_scale} \& \ref{Fig_scan}).

The overall mean plate scale turns out to be 1.213$\pm$0.0011\,$\arcsec$/px in the direction of Right Ascension and 1.209$\pm$0.00079\,$\arcsec$/px in Declination, respectively derived by maximum-likelihood fitting of the corresponding unbinned data sets with a Gaussian distribution).

\section{Conclusions}

The archiving work on the Schmidt plate collection of the AIU was done as part of a long-term project to support preservation and repeated usage of the world-wide fund of astronomical wide-field photographic plate collections. This international project strives for general on-line access by the scientific community to archives, plate metadata, and digitized plate images with optimal photometric and astrometric accuracy. The preparation of the AIU plate catalogue together with some statistical evaluation, as well as
the scanning procedure are described in this paper. The catalogue is included in the WFPDB, which is accessible online\footnotemark[1]. The database, among other things, provides information on the telescope, parameters of the observations, details from the logbook, and previews of the scanned Schmidt plates with limited resolution. Beyond that, the paper gives insights into the history of photographic observations carried out at the University Observatory Jena with the 60/90/180-cm Schmidt telescope between 1963 and 1982. 

An images of the digitized plates in their original file formats (FITS) are available upon request.

\acknowledgements We appreciate the cooperation with several of the observers, who carried out photographic plate observations, using the Schmidt camera of the University Observatory Jena. This research made use of tools provided by Astrometry.net. 
K.P.T. and M.K.T. acknowledge the support from the AIU  and the BG NSF DO-02-273/275 projects. 
 A.V.P. and V.V.H. acknowledge support by the German 
\emph{Deut\-sche For\-schungs\-ge\-mein\-schaft (DFG)\/} through project 
C7 of SFB/TR~7 ``Gravitationswellenastronomie''.


\begin{thebibliography}{}
\bibitem{Bar03} Barbieri, C., Blanco, C., Bucciarelli, B., Coluzzi, R., Di
  Paola,
  A., Lanteri, L., Li Causi, G.-L., Marilli, E., Massimino, P., Mezzalira, V.,
  Mottola, S., Nesci, R., Omizzolo, A., Pedichini, F., Rampazzi, F., Rossi, C.,
  Stagni, R., Tsvetkov, M., Viotti, R. 2003, Experimental Astronomy, 15.1, 29
\bibitem{ftools} Blackburn, J. K. 1995, in ASP Conf. Ser., Vol. 77,
  Astronomical Data Analysis Software and Systems IV, ed. R. A. Shaw,
  H. E. Payne, and J. J. E. Hayes (San Francisco: ASP), 367.
\bibitem{Br67} Breido, I. I.: 1967, Soviet Astron. 10, 714
\bibitem{Do68} Dorschner, J. and Friedemann, C.: 1968, AN 291, 7
\bibitem{Fr77} Friedemann, C., G\"urtler, J., Pfau, W.: 1977, AN 298, 327
\bibitem{HZ78} H\"ogner, W., Ziener, R.: 1978, AN 299, 159
\bibitem{Kue73} K\"uhn, L.: 1973, AN 294, 159
\bibitem{Mx71} Marx, S.: 1971, AN 292, 243
\bibitem{Mx79} Marx, S., Lehmann, H.: 1979, AN 300, 295
\bibitem{Mx65} Marx, S., Pfau, W.: 1965, Mitt. d. Univ.-Sternwarte Jena, No. 71
\bibitem{Mx66a} Marx, S., Pfau, W.: 1966a, Acta Astronomica 16, 81
\bibitem{Mx66b} Marx, S., Pfau, W.: 1966b, Die Sterne 42, 191
\bibitem{Mx67a} Marx, S., Pfau, W.: 1967a, AN 289, 290
\bibitem{Mx67b} Marx, S., Pfau, W.: 1967b, IBVS 206, 1
\bibitem{MxRi71} Marx, S., Pfau, W. Richter, N.: 1971, Jenaer Rundsch. 16, 294
\bibitem{MuCTK}  Mugrauer, M.: 2009, AN 330, 419
\bibitem{MuFiasco} Mugrauer, M., Avila, G.: 2009, AN 330, 430
\bibitem{MuSTK}  Mugrauer, M., Berthold, T.: 2010, AN 331, 449
\bibitem{MuFlechas} Mugrauer, M., Avila, G., Guirao, C.: 2014, AN submitted
\bibitem{Pf73} Pfau, W.: 1973, IBVS 787, 1
\bibitem{Pf79} Pfau, W.: 1979, AN 300, 21
\bibitem{Pf80} Pfau, W.: 1980, IBVS, 1874
\bibitem{PfF80} Pfau, W. and Friedemann, C.: 1980, AN 301, 69
\bibitem{Zw64} Zwicky, F.: 1964, AnAp 27, 300
\end{thebibliography}
\end{document}